\documentclass[aps,prd,reprint,groupedaddress]{revtex4-1}

\usepackage{amsmath,amssymb,amsfonts,graphicx,amscd,amsfonts,epsfig,color}

\newcommand {\beq} {\begin{equation}}
\newcommand {\eeq} {\end{equation}}
\newcommand {\beqa}{\begin{eqnarray}}
\newcommand {\eeqa}{\end{eqnarray}}

\begin{document}


\title{Complex Langevin calculations in finite density QCD
at large $\mu/T$\\
with the deformation technique}


\author{Keitaro Nagata}\email{k-nagata@kochi-u.ac.jp}
\affiliation{KEK Theory Center, 
High Energy Accelerator Research Organization,\\
1-1 Oho, Tsukuba, Ibaraki 305-0801, Japan}
\affiliation{Center of Medical Information Science, 
Kochi Medical School, Kochi University,\\
Kohasu, Oko-cho, Nankoku-shi, Kochi 783-8505, Japan}

\author{Jun Nishimura}\email{jnishi@post.kek.jp}
\affiliation{KEK Theory Center, 
High Energy Accelerator Research Organization,\\
1-1 Oho, Tsukuba, Ibaraki 305-0801, Japan}
\affiliation{Graduate University for Advanced Studies (SOKENDAI),\\
1-1 Oho, Tsukuba, Ibaraki 305-0801, Japan}

\author{Shinji Shimasaki}\email{shimasaki.s@gmail.com}
\affiliation{KEK Theory Center, 
High Energy Accelerator Research Organization,\\
1-1 Oho, Tsukuba, Ibaraki 305-0801, Japan}
\affiliation{Research and Education Center for Natural Sciences, Keio University,\\
Hiyoshi 4-1-1, Yokohama, Kanagawa 223-8521, Japan}


\date{\today}

\begin{abstract}
It is well known that investigating 
QCD at finite density by standard Monte Carlo methods
is extremely difficult due to the sign problem.
Some years ago, the complex Langevin method with gauge cooling
was shown to work at high temperature, i.e., in the deconfined phase.
The same method was also applied to QCD in the so-called 
heavy dense limit in the whole temperature region.
In this paper we attempt to apply this method to the 
large $\mu/T$ regime
with moderate quark mass
using four-flavor staggered fermions on a $4^3\times 8$ lattice.
While a straightforward application faces with the singular-drift problem,
which spoils the validity of the method, we overcome this problem
by the deformation technique proposed earlier.
Explicit results for the quark number density and the chiral condensate
obtained in this way for $3.2 \le \mu/T \le 5.6$
are compared with the results for the phase-quenched model
obtained by the standard rational hybrid Monte Carlo calculation.
This reveals a clear difference, which is qualitatively
consistent with the Silver Blaze phenomenon.
\end{abstract}


\maketitle


\section{Introduction}

The phase diagram of QCD at finite density and temperature
is speculated to have a very rich structure.
This is not only interesting from theoretical viewpoints
but also relevant to the physics related to heavy-ion collision experiments
and the interior structure of neutron stars.
%
However, the speculated phase structure still remains elusive
mainly because first-principle calculations based on lattice QCD
are extremely difficult at finite density
due to the complex fermion determinant,
which causes the so-called sign problem.

As a promising solution to this problem,
the complex Langevin method (CLM) \cite{Parisi:1984cs,Klauder:1983sp} 
has been attracting much attention recently.
In this method, based on the idea of stochastic 
quantization \cite{Parisi:1980ys,Damgaard:1987rr},
the expectation value of an observable 
is calculated using a stochastic process for \emph{complexified}
dynamical variables with the observable being extended holomorphically.
Since the method does not rely on the probabilistic interpretation 
of the Boltzmann weight, there is a chance to overcome 
the sign problem completely.

However, it is known that the method yields wrong results 
in some cases even if the stochastic process reaches equilibrium
without any problem.
This issue was discussed theoretically for the first time
in Refs.~\cite{Aarts:2009uq,Aarts:2011ax}
by considering the equality between 
the expectation value of an observable
defined by the stochastic process at each fictitious time
and the expectation value of the observable 
with respect to a complex weight, 
which satisfies the Fokker-Planck equation associated with the 
original theory.
If this equality holds,
the expectation value obtained by the stochastic process
in the long-time limit
gives the expectation value defined by the path integral formulation 
of the original theory \footnote{For this to be true,
it had long been considered
that all the eigenvalues of the Fokker-Planck Hamiltonian should have
a positive real part. However, 
it was argued in Ref.~\cite{Nishimura:2015pba} with explicit examples
that this condition is automatically satisfied if the 
aforementioned equality holds.}.
In proving this equality, a crucial role is played by 
the time-evolved observable, whose existence is 
implicitly assumed in Refs.~\cite{Aarts:2009uq,Aarts:2011ax}.
This is actually subtle and requires the condition that 
the probability distribution of the magnitude of the drift term 
in the stochastic process
should fall off exponentially or faster \cite{Nagata:2016vkn}.
On the other hand, if this condition is satisfied,
the integration by parts used in the argument can be justified. 
In this sense, one may regard the
above condition as a necessary and sufficient condition 
for justifying the CLM under such assumptions as the convergence
and the ergodicity of the stochastic process.
Roughly speaking, frequent appearance of large drifts during the 
stochastic process invalidates the CLM.
The validity of this criterion has been demonstrated in simple 
one-variable models \cite{Nagata:2016vkn}
and in semi-realistic models \cite{Nagata:2018net}.


There are actually two cases that can lead to the 
frequent appearance of large drifts.
One is the case in which the dynamical variables make frequent excursions
in the imaginary directions during the stochastic 
process, which is referred 
to as the excursion problem \cite{Aarts:2009uq,Aarts:2011ax}.
The other is the case in which the drift term has singularities
that are frequently visited during the stochastic 
process, which is referred to as the singular-drift 
problem \cite{Nishimura:2015pba}.
By avoiding these problems,
one can enlarge the validity region of the CLM.
For instance, 
the excursion problem can be solved
by the gauge cooling \cite{Seiler:2012wz}, 
which amounts to making a complexified gauge transformation after each
Langevin step in such a way that the imaginary part of the dynamical
variables is minimized \footnote{This technique was also used to
solve the singular-drift problem 
in the chiral Random Matrix Theory \cite{Nagata:2016alq}.
See Ref.~\cite{Bloch:2017sex}, however, for a case in which 
it does not work.}.
Theoretical justification of the gauge cooling has been 
given explicitly in Refs.~\cite{Nagata:2015uga,Nagata:2016vkn}.
On the other hand, Ref.~\cite{Ito:2016efb} proposed
to solve the singular-drift problem
by deforming the original system in such a way
that the dynamical variables keep away 
from the singularities of the drift term.
The results for the undeformed system
can be obtained
by extrapolating the deformation parameter to zero
using the parameter region in which the criterion for justifying the CLM
is satisfied.
This technique was applied successfully to matrix models
relevant to nonperturbative string theory \cite{Anagnostopoulos:2017gos}.
For other recent developments in the CLM, 
see Refs.~\cite{Makino:2015ooa,Tsutsui:2015tua,Fodor:2015doa,Hayata:2015lzj,Ichihara:2016uld,Aarts:2016qrv,Abe:2016hpd,Aarts:2016qhx,Salcedo:2016kyy,Bloch:2017ods,Aarts:2017vrv,Nishimura:2017vav,Doi:2017gmk,Fujii:2017oti,Basu:2018dtm}, for instance.

The gauge cooling made it possible
to apply the CLM to finite density QCD 
in the deconfined phase \cite{Sexty:2013ica,Fodor:2015doa}
and in the heavy dense limit \cite{Seiler:2012wz,Aarts:2013uxa,Aarts:2016qrv}.
In this paper, we attempt to investigate 
the large $\mu/T$ regime
with moderate quark mass
using a $4^3\times 8$ lattice \footnote{Preliminary results are presented 
in Lattice 2017 \cite{Nagata:2017pgc}. See also 
Refs.~\cite{Sinclair:2015kva,Sinclair:2016nbg,Sinclair:2017zhn} 
for related work.}.
The parameter region we are aiming at, however, is anticipated 
to be plagued by the singular-drift problem
according to the studies of the chiral Random Matrix 
Theory \cite{Mollgaard:2013qra}.
Indeed we encounter 
this problem and use the deformation technique to solve it.
By probing the probability distribution of the drift term,
we determine the region of the deformation parameter in which
the CLM is valid and make extrapolations to the undeformed model 
using the results within this region.
The baryon number density and the chiral condensate
thus obtained as a function of the quark chemical potential
are compared with
those obtained by the rational hybrid Monte Carlo (RHMC) calculation 
of the phase-quenched model,
which is defined by omitting the phase of the fermion determinant.
We observe a clear difference, which is qualitatively consistent with
the so-called Silver Blaze phenomenon in the full model.

This paper is organized as follows. 
In section \ref{sec:qcd-setup},
we briefly review lattice QCD at finite density.
In section \ref{sec:CLM_review},
we explain how we apply the CLM to finite density QCD
with
the gauge cooling and the deformation technique.
In section \ref{sec:results}, we present our results
and compare them with the results for the phase-quenched model.
Section \ref{sec:summary} is devoted to a summary and discussions.

\section{Lattice QCD at finite density}
\label{sec:qcd-setup}

Our calculation is based on lattice QCD on a 4D Euclidean periodic lattice
defined by the partition function
\begin{align}
  Z = \int \prod_{x\mu}dU_{x\mu} \, \mathrm{det}M(U,\mu) \ e^{-S_{\rm g}(U)} \ .
  \label{Z}
\end{align}
The dynamical variables $U_{x\mu}\in {\rm SU}(3)$ are the link variables,
where $x=(x_1,x_2,x_3,x_4)$ labels a site on the lattice 
and $\mu=1,2,3,4$ represents a direction with $\mu=1,2,3$ and $\mu=4$
being the spatial and temporal directions, respectively.
We work in units which set the lattice spacing to unity,
and denote the number of sites in the spatial and temporal directions
as $N_{\rm s}$ and $N_{\rm t}$, respectively.
The gauge action $S_{\rm g}(U)$ is given by
\begin{align}
  S_{\rm g}(U)=
-\frac{\beta}{6}\sum_{x}\sum_{\mu<\nu}
\mathrm{tr} \, (U_{x\mu\nu}+U_{x\mu\nu}^{-1}) \ ,
\end{align}
where 
the plaquette $U_{x\mu\nu}$ is defined 
by $U_{x\mu\nu}=U_{x\mu}U_{x+\hat\mu,\nu}U_{x+\hat\nu,\mu}^{-1}U_{x\nu}^{-1}$
with $\hat\mu$ being the unit vector in the $\mu$-direction.

In this work, we use the unimproved staggered fermions, 
for which the fermion matrix $M(U,\mu)$ in \eqref{Z} is given by
\begin{align}
  M(U,\mu)_{xy}&= m\delta_{xy}+\sum_{\nu=1}^{4}\frac{1}{2}\, \eta_\nu(x)
\left(e^{\mu \delta_{\nu 4}}U_{x\nu}\delta_{x+\hat\nu , y} \right. \nonumber \\
& \quad \quad \quad \quad \quad \quad 
 \left.  -e^{-\mu \delta_{\nu 4}}U_{x-\hat\nu , \nu}^{-1}
\delta_{x-\hat\nu ,  y}\right) \ ,
    \label{M}
\end{align}
where $\eta_\nu(x)=(-1)^{x_1+\cdots+x_{\nu-1}}$.
This represents four flavors of quarks 
with the degenerate quark mass $m$ and the quark chemical potential $\mu$. 
The quark field obeys the periodic/anti-periodic boundary conditions in the 
spatial/temporal directions, respectively.
Note that the fermion matrix $M(U,\mu)$ satisfies
$\epsilon_x M(U,\mu)_{xy}\epsilon_y=M(U,-\mu^*)^{\ast}_{yx}$
with $\epsilon_x=(-1)^{x_1+x_2+x_3+x_4}$ playing the role of $\gamma_5$.
Hence, for nonzero real $\mu$, 
the fermion determinant $\mathrm{det}M(U,\mu)$
becomes complex in general
causing the sign problem, which we overcome
by the CLM explained in the next section.

The observables we consider in this paper are the baryon number density
\begin{align}
  \langle n \rangle
  &=\frac{1}{3N_{\rm V}}\frac{\partial}{\partial \mu}\log Z  \ ,
  \label{baryon}
\end{align}
and the chiral condensate
\begin{align}
\langle  \Sigma \rangle
  &=\frac{1}{N_{\rm V}}\frac{\partial}{\partial m}\log Z \ ,
  \label{chiral}
\end{align}
where $N_{\rm V}=N_{\rm s}^3 N_{\rm t}$.
We use the standard noisy estimator to calculate these quantities.
Details related to this method are given in 
Appendix \ref{sec:noisy-estimator}.

\section{Complex Langevin method for finite density QCD}
\label{sec:CLM_review}

In this section, we explain how we apply 
the CLM \cite{Parisi:1984cs,Klauder:1983sp} 
to lattice QCD at finite density.
First we extend the link variables 
$U_{x\mu}\in {\rm SU}(3)$ to 
the complexified link variables 
$\mathcal U_{x\mu}\in {\rm SL}(3,\mathbb{C})$
and consider their fictitious time evolution 
based on the complex Langevin equation, which is given in its discrete form by
\begin{align}
&  \mathcal U_{x\mu}(t+\epsilon) \nonumber \\
&  =\exp\left(i\sum_{a=1}^{8}\lambda_a\left[-\epsilon v_{ax\mu}(\mathcal U(t))
    +\sqrt{\epsilon}\eta_{ax\mu}(t)\right]  \right) \ \mathcal U_{x\mu}(t) \ ,
    \label{CLE}
  \end{align}
where $t$ represents the discretized Langevin time with the stepsize $\epsilon$.
We have introduced the generators $\lambda_a$ $(a=1,\cdots,8)$ 
of the SU(3) algebra normalized 
as 
$\mathrm{tr} (\lambda_a\lambda_b) = \delta_{ab}$
and the real Gaussian noise $\eta_{ax\mu}(t)$ normalized as
$\langle \eta_{ax\mu}(t)\eta_{by\nu}(t')\rangle_\eta=
2\delta_{ab}\delta_{xy}\delta_{\mu\nu}\delta_{tt'}$,
where $\langle\cdots\rangle_\eta$ represents an average over $\eta$.
The drift term $v_{ax\mu}(\mathcal U)$ in \eqref{CLE}
is defined by 
analytic continuation of the one defined for $U_{x\mu}\in {\rm SU}(3)$ as
\begin{align}
  v_{ax\mu}(U)= D_{ax\mu} S(U) \equiv
\lim_{\varepsilon \to 0}
\frac{S(e^{i\varepsilon\lambda_a}U_{x\mu})-S(U_{x\mu})}{\varepsilon} \ ,
  \label{drift}
\end{align}
where $S(U)=S_{\rm g}(U)-\log \mathrm{det} M(U,\mu)$.

In order to calculate the VEV of a gauge-invariant observable $O(U)$,
we define $O(\mathcal U)$ by analytic continuation
and
its expectation value
\begin{align}
  \Phi(t) \equiv \langle O(\mathcal U(t))\rangle_{\eta}
\end{align}
in the CLM.
Then, under certain conditions, one can prove that
\begin{align}
  \lim_{t\to \infty}\lim_{\epsilon\to 0}\Phi(t)
=\frac{1}{Z}\int \prod_{x\mu}dU_{x\mu}O(U) \, e^{-S} \ ,
  \label{equality}
\end{align}
which implies that the left-hand side gives the VEV of $O(U)$ in the
original theory.

As is mentioned in the Introduction,
the proof of \eqref{equality} was first 
given in Refs.~\cite{Aarts:2009uq,Aarts:2011ax}
and was refined later by Ref.~\cite{Nagata:2016vkn}, which
showed that the necessary and sufficient condition 
for \eqref{equality} to hold
is that the probability distribution of the magnitude of
the drift term should fall off exponentially or faster.
In this work, we define the magnitude of the drift term as
\begin{align}
  u=\left(\frac{1}{8N_{\rm V}}\sum_{x\mu}\sum_{a=1}^{8}
|v_{ax\mu}(\mathcal U)|^2\right)^{\frac{1}{2}} \ ,
  \label{drift mag}
  \end{align}
and probe its probability distribution in order to see whether
the CLM is valid or not at each set of parameters.
The validity of this criterion is tested
not only in simple one-variable models \cite{Nagata:2016vkn}
but also in semi-realistic many-variable systems \cite{Nagata:2018net}.

It is known that the slow fall-off of the drift distribution,
which invalidates \eqref{equality},
is caused either by the excursion problem or by the singular-drift problem.
In finite density QCD, the former problem occurs
when the complexified link variables become far from unitary,
while the latter problem occurs when
the fermion matrix \eqref{M} has many eigenvalues close to zero
for the complexified link variables.

In order to solve the excursion problem, 
one can use
the gauge cooling, which
amounts to making a complexified gauge transformation 
after each Langevin step 
in such a way that the complexified link variables come
closer to a unitary configuration \cite{Seiler:2012wz}. 
The gauge transformation can be determined by
minimizing the unitarity norm
\begin{align}
  \mathcal N_u
  =\frac{1}{4N_{\rm V}}\sum_{x\mu}\mathrm{tr}
  \left[(\mathcal{U}_{x\mu})^\dagger 
\mathcal{U}_{x\mu}+(\mathcal{U}_{x\mu}^{-1})^\dagger \mathcal{U}^{-1}_{x\mu}
  -2\times \mathbf{1}_{3\times 3}\right] \ ,
\label{unitarity-norm}
\end{align}
which measures how far the link variables are from
a unitary configuration.
It has been shown explicitly \cite{Nagata:2015uga,Nagata:2016vkn}
that this additional procedure does not affect the argument for 
justifying the CLM.
The gauge cooling played a crucial role in enabling
the application of 
the CLM to finite density QCD 
in the deconfined phase \cite{Sexty:2013ica}
and in the heavy dense limit \cite{Seiler:2012wz,Aarts:2013uxa,Aarts:2016qrv}.

\begin{figure}[t]
\includegraphics{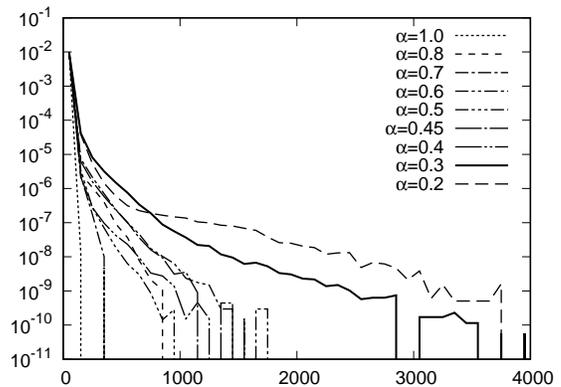}
\caption{\label{semilog_drift_mu0.70} The probability distribution of $u$,
the magnitude of the drift term defined by \eqref{drift mag},
is shown 
in a semi-log plot for $\mu=0.7$ with various $\alpha$. 
}
\end{figure}

At large $\mu/T$ with moderate quark mass,
the singular-drift problem occurs on top of the excursion problem.
In order to solve that problem,
we use 
the deformation technique \cite{Ito:2016efb},
which was applied successfully to matrix models relevant to 
superstring theory \cite{Anagnostopoulos:2017gos}.

In the case at hand,
we introduce a deformation parameter $\alpha\in \mathbb{R}$ 
in the fermion matrix \eqref{M} as
\begin{align}
M(U,\mu)_{xy}\to M(U,\mu)_{xy}+i\alpha\eta_4(x)\delta_{xy} \ .
  \end{align}
This deformation may be regarded as
adding an imaginary chemical potential
in the continuum theory.
Strictly speaking,
the extra term corresponds to adding a term
$i \alpha \bar{\psi}(x) (\gamma_4 \otimes \gamma_4) \psi(x)$
in the Lagrangian density of the continuum theory,
where the first $\gamma_4$ acts on
the spinor indices and the second $\gamma_4$ acts on the flavor indices.

For $\alpha$ large enough, 
the eigenvalue distribution of the fermion matrix develops 
a gap near the real axis,
which enables us to avoid the singular-drift problem.
When the
singular-drift problem occurs, the unitarity norm (\ref{unitarity-norm})
becomes large and it sometimes becomes uncontrollable. This problem
is cured when the singular-drift problem is avoided by sufficiently
large $\alpha$.

We probe the drift distribution at each $\alpha$, and determine the
range of $\alpha$ for which the obtained results are reliable.
Extrapolating the results within this range of $\alpha$
to $\alpha=0$, we obtain the results
for the original theory.
Considering the symmetry of the deformed theory 
under $\alpha\leftrightarrow -\alpha$,
we choose the fitting function 
to be a linear function of $\alpha^2$.


\begin{figure}[t]
\includegraphics{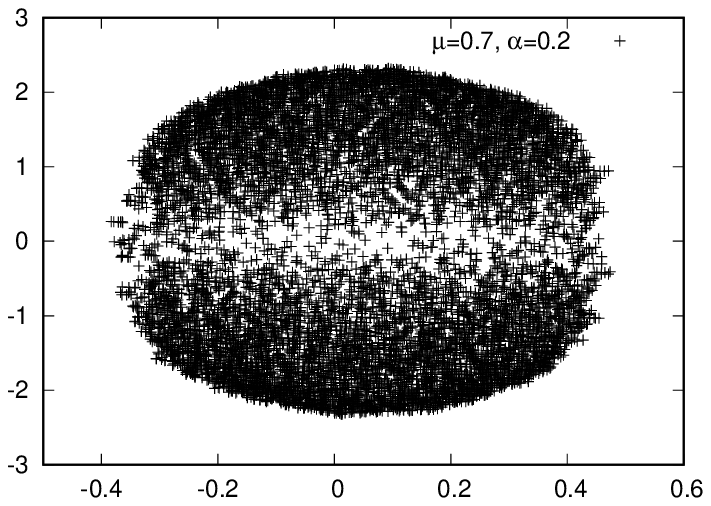}
\includegraphics{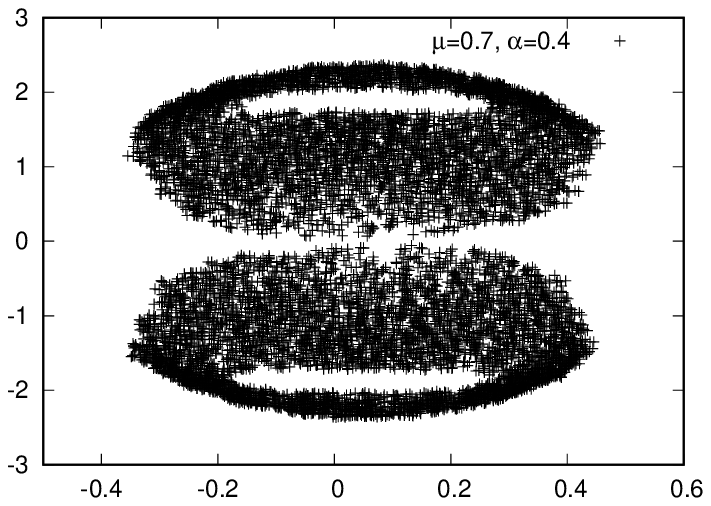}
\caption{\label{semilog_drift_mu0.70-2} The eigenvalue distribution 
of the fermion matrix 
is shown for $\mu=0.7$ with $\alpha=0.2$ (Top) and $\alpha=0.4$ (Bottom).}
\end{figure}

\section{Results}
\label{sec:results}

In this section, we show our results 
for finite density QCD obtained by the CLM 
as explained in the previous section.
We use a $4^3\times 8$ lattice 
with the gauge coupling $\beta=5.7$
and the quark mass $m=0.05$.
The quark chemical potential $\mu$ is taken to be $0.4 \le \mu \le 0.7$,
which implies that the physical $\mu/T$ ranges from $3.2$ to $5.6$.
The Langevin process \eqref{CLE} is performed for 
the total Langevin time $50\sim 150$ 
with a fixed stepsize $\epsilon=10^{-4}$.
We present results for the baryon number density \eqref{baryon} and 
the chiral condensate \eqref{chiral},
which are compared with
those for the phase-quenched model
obtained by the standard RHMC calculation.

First we check the validity of the CLM
by probing the probability distribution of the drift term, 
which is shown in Fig.~\ref{semilog_drift_mu0.70} for $\mu=0.7$ 
with various $\alpha$.
We find that the probability distribution 
falls off exponentially or faster 
for $\alpha\geq 0.4$, 
while a power-law tail develops for $\alpha=0.2, 0.3$. 
This implies that the CLM is valid for $\alpha \geq 0.4$ at $\mu=0.7$.

The power-law tail of the probability distribution 
for $\alpha\lesssim 0.3$ is actually
due to the singular-drift problem caused 
by near-zero eigenvalues of the fermion matrix
as one can see from Fig.~\ref{semilog_drift_mu0.70-2}.
Indeed there are many eigenvalues distributed around the 
origin for $\alpha=0.2$, which is not the case
for $\alpha=0.4$ owing to the gap developing along the real axis.

\begin{figure}[t]
\includegraphics{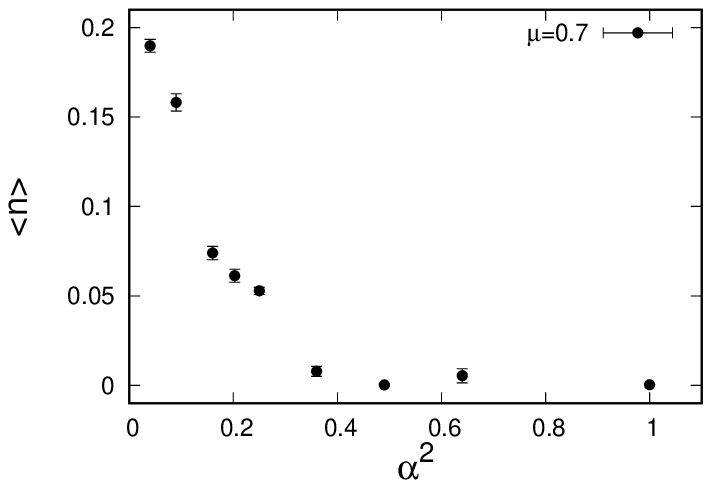}
\includegraphics{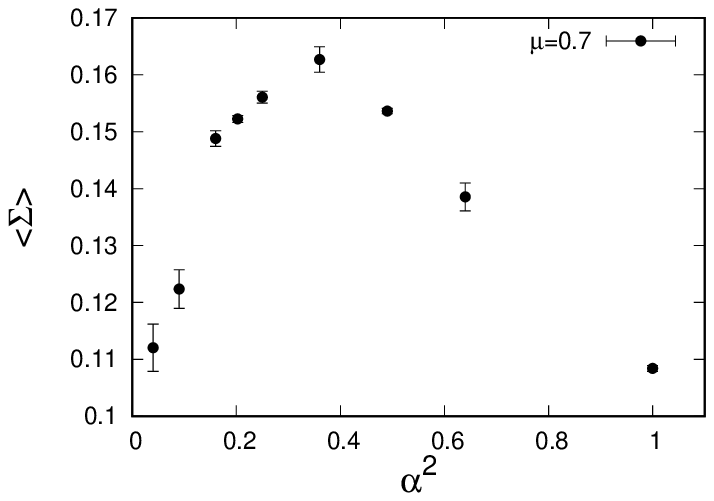}
\caption{\label{mu0.70_alpha} The baryon number density (Top) and 
the chiral condensate (Bottom) obtained by the CLM 
are plotted against $\alpha^2$ for $\mu=0.7$. 
}
\end{figure}

\begin{figure}[t]
\includegraphics{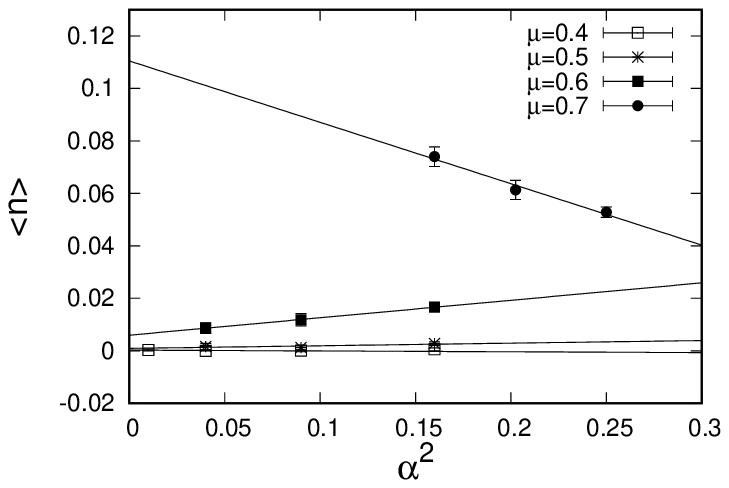}
\includegraphics{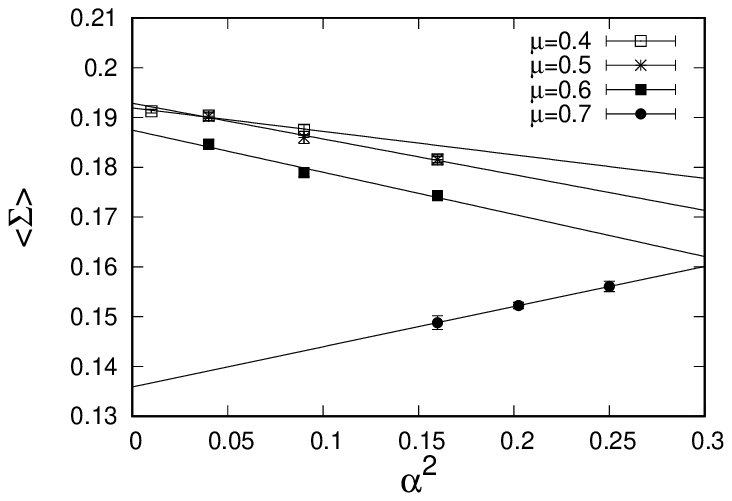}
\caption{\label{extrapolation} The baryon number density (Top)
and the chiral condensate (Bottom) obtained by the CLM are plotted against 
$\alpha^2$ for $\mu=0.4, 0.5, 0.6, 0.7$.
We present only the data points that are reliable
in the light of the drift distribution.
The straight lines represent linear extrapolations
to $\alpha = 0$ with respect to $\alpha^2$.}
\end{figure}

In Fig.~\ref{mu0.70_alpha}, we plot the baryon number density 
$\langle n \rangle$ (Top)
and 
the chiral condensate 
$\langle \Sigma \rangle$ (Bottom)
obtained by the CLM against $\alpha^2$ for $\mu=0.7$. 
Note that the data points for $\alpha\lesssim 0.3$ should be discarded
since the CLM is not valid there.
We find that $\langle n \rangle$ drops to zero for $\alpha\geq 0.6$
and that $\langle \Sigma \rangle$ changes its behavior at $\alpha \sim 0.6$.
These observations suggest
the existence of a phase transition
at $\alpha\sim0.6$.
Thus we are led to use only the data points for $\alpha=0.4, 0.45, 0.5$ 
for the extrapolation to $\alpha = 0$ for $\mu=0.7$.
These are shown 
in Fig.~\ref{extrapolation} by circles,
which can be fitted to a straight line.
In the same figure, we also plot
the reliable data points 
obtained for other values of $\mu$
together with the linear extrapolation using
$\alpha=0.1, 0.2, 0.3$ for $\mu=0.4$ and
$\alpha=0.2, 0.3, 0.4$ for $\mu=0.5, 0.6$.

When we performed 
complex Langevin simulations
\emph{without} deformation,
the history of observables typically show
occasional spikes,
which makes it difficult to reduce the statistical error
within a reasonable computing time \cite{Nagata:2016mmh}.
This problem does not occur in all the cases investigated here
with the deformation.


\begin{figure}[t]
\includegraphics[width=7.5cm]{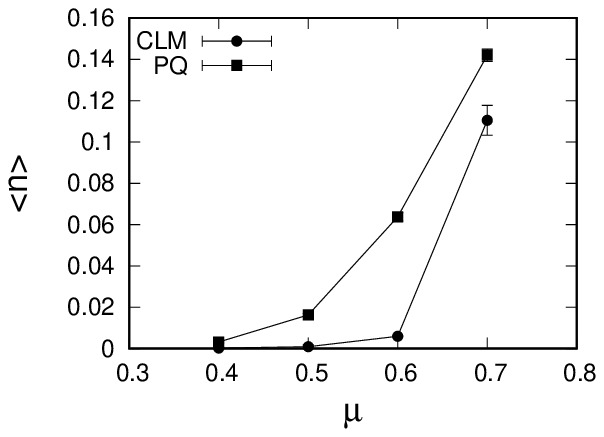}
\includegraphics[width=7.5cm]{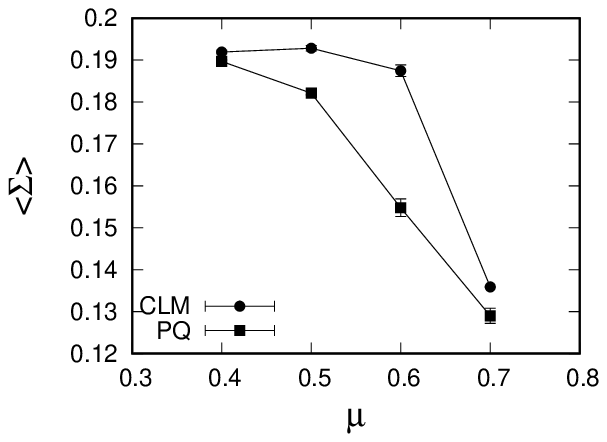}
\caption{\label{baryon and chiral} The extrapolated values 
of the baryon number density (Top)
and the chiral condensate (Bottom) obtained from Fig.~\ref{extrapolation}
are plotted against $\mu$. The solid lines are drawn to guide the eye.}
\end{figure}

In Fig.~\ref{baryon and chiral}, 
the extrapolated values for the baryon number density (Top)
and the chiral condensate (Bottom) are plotted against $\mu$ by circles.
In the same figure, we also
plot by squares 
the results for the phase-quenched model 
obtained by the standard RHMC calculation,
which reveals a clear difference.
The results of the CLM
show that 
the baryon number density is almost zero for $\mu\lesssim 0.6$ 
and has a sharp increase within $0.6\lesssim \mu \lesssim 0.7$.
Correspondingly, 
the chiral condensate is almost constant for $\mu\lesssim 0.6$,
and starts to decrease rapidly within $0.6\lesssim \mu \lesssim 0.7$.
While certain systematic errors due to 
the $\alpha\to 0$ extrapolation are considered to exist,
the rapid change within $0.6\lesssim \mu \lesssim 0.7$
should be robust judging 
from the qualitative difference
of the $\alpha$ dependence for $\mu=0.6$ and $\mu=0.7$
seen in Fig.~\ref{extrapolation}.
The results for the phase-quenched model, on the other hand,
show a milder $\mu$ dependence.
The onset of the baryon number density occurs 
around $\mu\sim 0.4$, 
where the chiral condensate starts to decrease.

Note that the value of $\beta$ is chosen to be large ($\beta=5.7$)
in order to avoid the excursion problem that occurs
at smaller $\beta$ similarly to the situation 
found in Ref.~\cite{Fodor:2015doa}.
Because of this,
the corresponding lattice spacing is well below $0.1~{\rm fm}$ according to
a crude extrapolation from the data obtained in Ref.~\cite{Fodor:2015doa}.
Considering that our lattice is $4^3\times 8$,
this implies that
the temperature is quite high, whereas
the physical volume of the spatial lattice is 
much smaller than the QCD scale.
In fact, despite the high temperature, 
the Polyakov line vanishes for $\mu \le 0.6$
as we show in Appendix \ref{sec:polyakov-line}.
This can be understood as a consequence of the finite spatial volume
effects since
the increase of free energy for having one quark in such a small spatial region
is much larger than that in the infinite volume at the same temperature.
In other words,
the temperature is actually ``low'' compared with the scale of the spatial 
directions due to the chosen aspect ratio.
It is therefore not so surprising that
the observables behave more like those at low temperature
in a usual setup.

In full QCD at zero temperature in the infinite volume limit, 
physical observables are independent 
of $\mu$
up to $\mu\sim m_{\mathrm{N}}/3$
with $m_\mathrm{N}$ being the nucleon mass.
On the other hand,
in the case of the phase-quenched model, physical observables 
are independent of $\mu$
up to $\mu\sim m_{\pi}/2 (<m_\mathrm{N}/3)$
with $m_\pi$ being the pion mass.
The $\mu$ independence of full QCD 
within the region $m_{\pi}/2 <\mu< m_\mathrm{N}/3$,
which is commonly referred to as the ``Silver Blaze phenomenon''
in the literature,
is expected to occur due to the effect
of the phase of the complex fermion determinant.
Our results are qualitatively consistent with this expectation.



\section{Summary and discussions}
\label{sec:summary}

We have made an attempt to extend the success of 
the CLM in investigating finite density QCD 
in the deconfined phase or in the heavy dense limit
to the large $\mu/T$ regime
with moderate quark mass.
In this exploratory work, we use a $4^3\times 8$ lattice 
with four-flavor staggered fermions
and calculate the baryon number density and the chiral condensate 
as a function of the quark chemical potential.
The reliability of the obtained results is judged by 
the probability distribution of the magnitude of the drift term.
As in the previous work, the excursion problem is avoided
by the gauge cooling.
In addition to this, the singular-drift problem has to be overcome
in the parameter regime we explore.
The deformation technique,
which was shown to be useful in the case of matrix models 
for superstring theory,
turns out to be useful also in the present case.
By probing the probability distribution of the magnitude of the drift term,
we find that the singular-drift problem can be cured and reliable data
can be obtained unless the deformation parameter is too small.
The results for the original theory are obtained by extrapolation
using only the reliable data.
Thus we are able to obtain explicit results
in the region $3.2 \leq \mu/T\leq 5.6$ with moderate quark mass.

By comparing the results of the CLM with those obtained 
by the RHMC calculations in the phase-quenched model,
we observe that the onset of the 
baryon number density in the full model occurs at larger $\mu$
than in the phase-quenched model, which is
qualitatively consistent with the Silver Blaze phenomenon,
which occurs at zero temperature in the infinite volume.
In order to confirm this phenomenon,
we clearly need to increase the lattice size.
We have already started simulations
on a $8^3\times 16$ lattice \cite{workinprog}
and found that the CLM actually gives correct results 
even in the region of relatively large chemical potential 
without the deformation technique.
Preliminary results 
for the baryon number density and the chiral condensate 
show a rapid change twice as we increase the chemical potential,
which may be interpreted as the phase transitions to 
the nuclear matter and to the quark matter.
We hope to report on these results in the forthcoming publication.


\section*{Acknowledgements}

The authors would like to thank 
Y.~Ito, T.~Kaneko,
H.~Matsufuru, K.~Moritake, A.~Tsuchiya and S.~Tsutsui
for valuable discussions.
Computations were carried out on 
Cray XC40 at YITP in Kyoto University,
SX-ACE at CMC and RCNP in Osaka University
and PC clusters at KEK.
K.~N.\ and J.~N.\ were supported in part by Grant-in-Aid 
for Scientific Research (No.\ 26800154 and 16H03988, respectively) 
from Japan Society for the Promotion of Science. 
S.~S.\ was supported by the MEXT-Supported Program 
for the Strategic Research Foundation at Private Universities 
``Topological Science'' (Grant No. S1511006).

\appendix

\section{Details of the noisy estimator}
\label{sec:noisy-estimator}

In order to calculate the drift term
\begin{align}
v_{a x \mu}^{\rm (f)}   &=
- \mathrm{tr} \, ( M^{-1} D_{ax \mu} M) 
%
  \label{fermion-drift}
\end{align}
obtained from the fermion determinant,
we use the standard noisy estimator.
In this section, we present the details of how we use this method
in our calculation.

The idea is to replace the trace in (\ref{fermion-drift}) by
\begin{align}
v_{a x \mu}^{\rm (f)}   &=
- \varphi^{*} 
 M^{-1} D_{ax \mu} M \varphi 
= - \psi^{*}  D_{ax \mu} M \varphi \ ,
  \label{fermion-drift-noisy}
\end{align}
where $\varphi$ is a random complex vector generated with the normalized
Gaussian distribution. The other complex vector $\psi = (M^\dag)^{-1} \varphi$ 
in (\ref{fermion-drift-noisy})
can be calculated from $\varphi$ as $\psi = M \chi$,
where $\chi$ is obtained by solving $M^\dag M \chi = \varphi$ 
using the conjugate gradient method.

The above procedure is exact
if we take an average over infinitely many $\varphi$ generated randomly. 
In practice, we generate the random vector only once at each Langevin step
and estimate the trace using it.
The use of this approximation does not yield any 
systematic errors in the CLM in the stepsize $\epsilon \rightarrow 0$ 
limit since the associated Fokker-Planck equation 
remains the same \cite{Batrouni:1985jn}.
When we evaluate the trace involved in the observables
(\ref{baryon}) and (\ref{chiral}),
we take an average over twenty $\varphi$'s generated randomly.

See Refs.~\cite{Sinclair:2015kva,Sinclair:2016nbg,Sinclair:2017zhn}
for a more sophisticated way to implement the noisy estimator,
which ensures that the drift term becomes real 
for unitary link variables at $\mu=0$.

\section{The results for the Polyakov line}
\label{sec:polyakov-line}

\begin{figure}[t]
\includegraphics{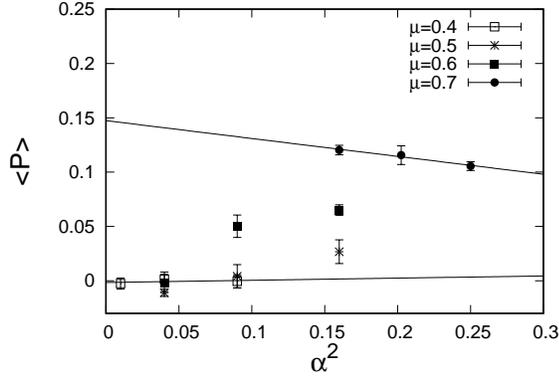}
\caption{\label{pol-extrapolation} The Polyakov line 
obtained by the CLM is plotted against 
$\alpha^2$ for $\mu=0.4, 0.5, 0.6, 0.7$.
We present only the data points that are reliable
in the light of the drift distribution.
The straight lines represent linear extrapolations
to $\alpha = 0$ with respect to $\alpha^2$ for $\mu=0.4$ and 
$\mu=0.7$.}
\end{figure}

In this section we present our results for the Polyakov line
defined by
\begin{align}
P  &=\frac{1}{3 N_{\rm s}^3}
\sum_{\vec{x}}
\mathrm{tr} \, (U_{(\vec{x},0) 4} U_{(\vec{x},1) 4} \cdots 
U_{(\vec{x},N_{\rm t}-1) 4}) \ .
  \label{polyakov}
\end{align}
Figure \ref{pol-extrapolation} shows the 
expectation value $\langle P \rangle$ plotted against
$\alpha^2$. 
We plot only the reliable data from the viewpoint
of our criterion based on the drift distribution.
For $\mu=0.4$ and $\mu=0.7$, we can fit the data
to a straight line.
For $\mu=0.5$ and $\mu=0.6$, we find that the data are close to zero 
for sufficiently small $\alpha$.

In the phase-quenched model, the expectation value
$\langle P \rangle$ is obtained
as $0.594(1)$, $0.749(2)$, $0.121(3)$, $0.146(4)$
for $\mu=0.4$, 0.5, 0.6, 0.7, respectively.

\bibliography{clmqcd_ref}

\end{document}